\documentclass[english]{iopart}
\usepackage[T1]{fontenc}
\usepackage[latin9]{inputenc}
\usepackage{graphicx}

\makeatletter

\providecommand{\tabularnewline}{\\}

\usepackage{iopams}
\usepackage{setstack}

\makeatother

\usepackage{babel}

\begin{document}

\paper{Conjecture on the maximum cut and bisection width in random regular
graphs}

\author{Lenka Zdeborov\'a$^{1}$, Stefan Boettcher$^{2}$}

\address{$^{1}$Theoretical Division and Center for Nonlinear Studies, Los
Alamos National Laboratory, NM 87545 USA}

\address{$^{2}$Physics Department, Emory University, Atlanta, Georgia 30322;
USA, }

\ead{lenka.zdeborova@gmail.com, sboettc@emory.edu}

\begin{abstract}
Asymptotic properties of random regular graphs are object of extensive
study in mathematics. In this note we argue, based on theory of spin
glasses, that in random regular graphs the maximum cut size asymptotically
equals the number of edges in the graph minus the minimum bisection
size. Maximum cut and minimal bisection are two famous NP-complete
problems with no known general relation between them, hence our conjecture
is a surprising property of random regular graphs. We further support
the conjecture with numerical simulations. A rigorous proof of this
relation is obviously a challenge. 
\end{abstract}

\section{Introduction}

Maximum cut and minimal bisection are two famous problems in graph
theory. Given a graph, i.~e. a set of nodes $V$ and a set of edges
$E$, the goal in the maximum cut problem is to split the set of nodes
into two groups in such a way that the number of edges connecting
the two groups is the largest possible. In the minimal bisection problem
the goal is to split the set of nodes into two equally sized groups
in such a way that the number of edges between the two groups is the
smallest possible. Minimal bisection is also known under the name
of graph bi-partitioning. Both these problems are recognized as NP-complete
\cite{GareyJohnson79}, and both have a large number of applications
in computer science and engineering. Some intensively studied applications
of the graph partitioning problem are circuits design \cite{A+K},
or data clustering and load balancing in parallel computing \cite{HL}.
For applications of the max-cut problem see, e.~g.,  the survey article
in Ref. \cite{PoljakTuza95}.

Random $r$-regular graphs are randomly chosen from all those graphs
having $N$ nodes and the degree of each node fixed to $r$. Determining
the asymptotic size of their max-cut or their min-bisection (bisection
width) are classical problems in random graph theory, see Refs.~\cite{Bollobas88,KostochkaMelnikov92,MonienPreis01,DiazDo03,BertoniCampadelli06,DiazSerna07}
for the best known lower and upper bounds. However, as far as we know,
no explicit relation between the max-cut size and bisection width
is known in the graph theoretical literature. An exception is provided in
Ref.~\cite{DiazSerna07}, where the same approximative algorithm is used
to provide an upper bound for bisection width and lower bound for
max-cut.

The main purpose of this note is to conjecture that in random regular
graphs the size of the max-cut is asymptotically equal to the number
of edges minus the size of the min-bisection. Our conjecture states
that in the large $N$ limit, 
\begin{equation}
|MC|=|E|-|BW|+o(|BW|)\,,
\label{con}
\end{equation}
where $|E|$ is the total number of edges. 

\begin{figure}
\begin{centering}
\includegraphics[width=0.6\linewidth]{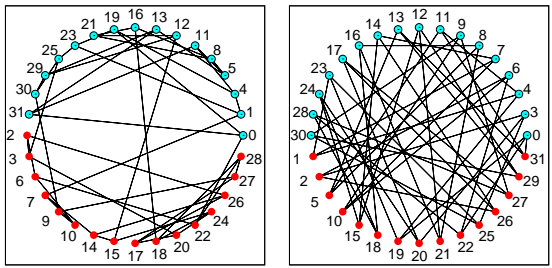} 
\par\end{centering}

\caption{\label{fig:example}Two different drawings of the same
randomly generated 3-regular graph with $N=32$ nodes. Left: Example
of a minimal bisection of the graph; only $6$ edges are present between
the group of blue (up) and red (down) nodes. Right: Example of a maximum
cut, only $5$ edges are present between two nodes of the same color.}

\end{figure}

In Fig.~\ref{fig:example}, we present two different drawings of
the same 3-regular graph with $N=32$ nodes. The left side is a minimal
bisection of size $|BW|=6$, the right side shows a maximum cut of
size $|MC|=rN/2-5$. It illustrates that Eq.~(\ref{con}) is a highly
non-intuitive result, since on a given graph there is no straightforward
relation between the set of edges in the maximum cut and the minimal bisection.
As we will show in the rest of this note, hints of this conjecture
already appeared in various forms in the spin glass literature. Our
goal is to collect arguments for its justification, state them in
a language that does not require knowledge of the replica or cavity
computations, provide evidence from precise numerical simulations,
and most importantly, clarify the conditions under which this conjecture
holds and discuss its generalizations.

\section{Statistical physics formulation of the problems}

In statistical physics, max-cut and bi-partitioning can be formulated in terms
of finding the ground state of an Ising model on random $r$-regular
graphs. For any graph, the general Ising model Hamiltonian reads 
\begin{equation}
{\cal H}=-\sum_{(ij)\in E}J_{ij}S_{i}S_{j}\,,
\label{Ham}
\end{equation}
where the sum extends over all edges in the graph, $J_{ij}$ is the
interaction strength, and $S_{i}\in\{-1,+1\}$ are the Ising spin
variables. The max-cut problem is cast as a ground state of the anti-ferromagnetic
Ising model, i.~e. minimization of (\ref{Ham}) with $J_{ij}=-1$ for
all $(ij)\in E$ with respect to the values of the spins $\{S_{i}\}$.
The min-bisection, or graph bi-partitioning, is a ground state of
the ferromagnetic Ising model with magnetization fixed to zero, i.~e.
minimization of (\ref{Ham}) with $J_{ij}=1$ for all $(ij)\in E$
subject to a constraint $\sum_{i}S_{i}=0$. Let $E_{{\rm GS}}(\{J_{ij}\})$
be the energy of the corresponding ground state, then the size of
the bisection width and the max-cut are 
\begin{equation}
|{\rm BW}|=\frac{|E|+E_{{\rm GS}}(\{J_{ij}\})}{2}\,,\quad|{\rm MC}|=
\frac{|E|-E_{{\rm GS}}(\{J_{ij}\})}{2}\,, \label{mcmc}
\end{equation}
where $|E|$ is again the total number of edges.

One can interpolate between the max-cut and min-bisection problems
by taking the interactions $J_{ij}$ uniformly at random from a bimodal
distribution 
\begin{equation}
P(J_{ij})=\rho\delta(J_{ij}+1)+(1-\rho)\delta(J_{ij}-1)
\end{equation}
and by fixing the magnetization to zero when needed. The disorder
in the interactions induces frustration on any kind of loopy lattice,
in which case the Hamiltonian (\ref{Ham}) then provides a model for
a spin glass \cite{MezardParisi87b}. For sparse random regular graphs the conjecture discussed here can be generalized as: The ground state energy
of (\ref{Ham}) is asymptotically independent of $\rho$, i.~e. in
the large $N$ limit it is
\begin{equation}
E_{{\rm GS}}(\{J_{ij}\},\rho)=Ne_{{\rm
    GS}}+o(N),\qquad(N\to\infty,\,0\leq\rho\leq1).  \label{gen}
\end{equation}

\section{Previous results}

In statistical physics of disordered systems, the replica or
cavity method \cite{MezardParisi87b,MezardParisi01} and a replica
symmetry breaking scheme \cite{Parisi80b}  is used to compute the exact ground
state of Hamiltonian (\ref{Ham}) at zero magnetization. Unfortunately,
these techniques are not rigorous, although in many models their result
have been proven, see e.~g. Refs.~\cite{Talagrand06,DemboMontanari08}.

Using the replica method, Fu and Anderson \cite{FuAnderson86} computed
the ground state energy of graph bi-partitioning ($\rho=0$) on dense
random graphs, i.~e. when the degree $r=pN$ ($0<p\le1$) is a constant independent of the graph size $N$. Their result reads 
\begin{equation}
E_{{\rm GS}}^{{\rm    dense}}=U_{SK}N^{\frac{3}{2}}\sqrt{p(1-p)}+o(N^{\frac{3}{2}}),
\label{FuA}
\end{equation}
where $U_{SK}$ is the ground state energy density of the Sherrington-Kirkpatrick
model \cite{SherringtonKirkpatrick75} computed by the Parisi formula \cite{Parisi80b}, a numerical
evaluation giving $U_{SK}=-0.763219\ldots$. They obtained this result
by realizing that on the dense graphs the replica equations for Hamiltonian
(\ref{Ham}) at zero magnetization are basically identical to the
replica equations for the Sherrington-Kirkpatrick model. Moreover, the
minimal bisection of a graph $G$ plus the maximum-cut of the
complement of $G$ (i.~e., the graph composed of edges that are not present in $G$) equals $(N/2)^2$. Using this identity plus Eqs.~(\ref{mcmc}) and (\ref{FuA}), we obtain that the bisection width is the number of edges minus the size of the max-cut $|{\rm BW}|=pN^{2}/2-|{\rm MC}|+o(N^{3/2})$.

Similarly, using results on the ground state energy of the spin glass, $\rho=1/2$, Refs.~\cite{MezardParisi87c,SherringtonWong87,WongSherrington87,WongSherrington88}
computed the bisection width on sparse random regular graphs. 
In the sparse case, the size of the bisection width is linear in the size of the system and one therefore obtains
$|{\rm BW}|=|E|-|{\rm MC}|+o(N)$. This relation can be proven rigorously
on sparse random graphs with large degree in the first two orders in the degree, in particular, $|BW|=rN/2+U_{SK}N\sqrt{r}+o(\sqrt{r})+o(N)$ \cite{Montanari09}.

However, the existing literature never discusses for what ensembles of random graphs the above results hold. A counter-example is provided by the Erd\H{o}s-R\'enyi random graphs, where every edge is present with probability $\alpha/(N-1)$. On the Erd\H{o}s-R\'enyi random graphs the spin glass model, $\rho=1/2$, and the max-cut, $\rho=1$, have a positive ground state energy above the percolation threshold, $\alpha>1$. In contrast, the bisection
width, $\rho=0$, of an Erd\H{o}s-R\'enyi graph is positive only above
$\alpha=2\ln{2}$, at which point the giant component reaches size
$N/2$. Thus, we need to discuss the theoretical arguments for sparse random graphs and specify the conditions under which the max-cut and bisection width are related.

Note also that on dense graphs, although conjecture (\ref{con})
holds, the ground state energy of (\ref{Ham}) is not
$\rho$-independent. Hence, generalization (\ref{gen}) holds only on sparse graphs. E.~g., for $p=1$ and $\rho=0$ the ground state energy is zero, whereas for $\rho=1/2$ the ground state energy is $U_{SK}N^{3/2}$.

For completeness, let us note that the model (\ref{Ham}) on
random graphs without the condition on zero magnetization was studied
in Ref.~\cite{CastellaniKrzakala05}. It was found that there is an
$r$-dependent critical value of $0<\rho_{c}(r)<1/2$ such that for
$\rho>\rho_{c}(r)$ the model with zero magnetization or non-fixed
magnetization are asymptotically equivalent, and for $\rho<\rho_{c}(r)$ the two
are different as the second develops a non-zero magnetization. From
the one-step replica symmetry breaking solution, Ref.~\cite{CastellaniKrzakala05} found, e.~g., $\rho_{c}(3)=0.142$ and $\rho_{c}(r)\to1/2$ for $r\to\infty$.

\section{Theoretical arguments}

We will argue that on sparse random graphs the ground state of
(\ref{Ham}) at zero magnetization does not depend on the fraction of
anti-ferromagnetic bonds $\rho$. The main part of the argument is
based on the fact that sparse random
graphs (with finite mean of the degree distribution) are locally tree
like, i.~e. the length of the shortest cycle passing trough a random
node diverges when $N\to\infty$. On a tree, all dependence on $\rho$
can be \textquotedbl{}pushed\textquotedbl{} to the boundary conditions
by using recursively from the root the  gauge transformations $J_{ij}\to\sigma_{i}J_{ij}\sigma_{j}$
and $s_{i}\to\sigma_{i}s_{i}$, where $\sigma_{i}\in\{\pm1\}$ are
chosen in such a way as to yield, say, $J_{ij}=-1$
for all $(ij)$\footnote{Any other pre-defined configuration of
  $J_{ij}$'s would do as well.}. 

Note that this gauge transformation always conserves
the Hamiltonian (\ref{Ham}). So we \textquotedbl{}only\textquotedbl{} need to discuss the dependence on the boundary conditions. There is a set of properties on the boundary conditions that do influence behavior in the bulk of the tree, let us call these the {\it relevant} properties. To make the connection between the system on trees and on the random graph, we need to consider boundary conditions on the tree having the same {\it relevant} properties as a configuration taken uniformly at random from the zero-temperature Boltzmann measure associated with (\ref{Ham}) on the random graph. 

Moreover, in order to argue in favor of our conjecture (\ref{gen}), we need the {\it relevant} properties of the boundary conditions to stay unchanged after the gauge transformation. A particular {\it relevant} property is the total magnetization on the boundary conditions. Note that gauge-flipping of any finite fraction of random bonds in the tree causes a random half of the spins on the boundary conditions to flip. Hence, only the zero value of magnetization can be treated this way. This is a first important limitation of the conjecture (\ref{gen}) -- the $\rho$-independence of the ground state of (\ref{Ham}) at zero magnetization does not generalize to non-zero values.

We said that all the {\it relevant} properties, not only the
magnetization, of the boundary conditions need to be conserved by the
gauge transform. This entails an impasse in the mathematical rigor of
our discussion and we have to resort to non-rigorous arguments
implicit in the cavity method. In the cavity method approach M\'ezard
and Parisi \cite{MezardParisi01} argue that the space of
configurations and boundary conditions can be split into states. Every
state has an associated set of boundary conditions in such a way that
within each there is no dependence of the bulk properties on the
precise boundary conditions corresponding to the state. This notion
is familiar from the Ising ferromagnet in the low temperature phase,
where there are two such states.  M\'ezard and Parisi \cite{MezardParisi01} treat the case where the number of states grows exponentially with the size of the system, this is called replica symmetry breaking. If there is independence of the bulk on the boundary conditions, then there are no {\it relevant} properties, and an empty set is certainly conserved by the above gauge transformation. Thus, to finish our argument we \textquotedbl{}only\textquotedbl{} need to show that the solution of the cavity equations (that describe the splitting into states) is $\rho$-independent or, in other words, conserved by the gauge transformation. 

The cavity equations are written in terms of local magnetic fields
and their distributions over the graph edges (and over the different
states, if the corresponding problem is glassy)\footnote{In computer science
the local magnetic fields are known as the beliefs in the belief
propagation algorithm, and different states correspond to different
belief propagation fixed points.}. It follows from the cavity equations
that, if there is a global symmetry between positive and negative
fields, then the system has to have zero magnetization. However, the
opposite is not true: requirement of zero magnetization does now imply
the distribution of fields to be symmetric around zero. The
inhomogeneity in the graph degree may lead to zero magnetization
without overall plus-minus symmetry, as can be illustrated again by
the example of Erd\H{o}s-R\'enyi graphs above the percolation
threshold, see Ref.~\cite{SulcZdeborova09}. There, at $\rho=0$ the denser parts of the
graph are more likely to have positive (or negative) fields, and sparser parts have excessive
negative (or positive) fields. Again, such a non-trivial symmetry
breaking is not conserved by the gauge transform and, hence, on
Erd\H{o}s-R\'enyi graphs the ground state of (\ref{Ham}) may be (and in fact is) $\rho$-dependent. Other ensembles of non-regular random graphs may also have this degree-fluctuations driven symmetry breaking and hence no reason for validity of (\ref{gen}), in particular for low values of $\rho$ where the equilibrium value of magnetization is not zero\footnote{On the other hand the $\rho$-independence may be valid for $\rho$ above some critical value. For example, the spin glass $\rho=1/2$ is equivalent to the anti-ferromagnet $\rho=1$ on other ensembles of sparse random graphs where the bisection $\rho=0$ is not.}.

Random regular graphs, on the other hand, have no inhomogeneity
in degree and they locally look the same from any node in the graph.
Moreover, for $\rho=0$ and $\rho=1$ there is no inhomogeneity in
the interactions $J_{ij}$ either, hence, the cavity fields (or their
distribution over states) have to be the same on every edge.
In such a case, the only way to obtain zero magnetization is to have
cavity-field distributions symmetric around zero. Hence one obtains
the same cavity equations for both, the graph bisection ($\rho=0$)
and the max-cut ($\rho=1$) problems. For the remaining values of
$0<\rho<1$, the neighborhood of every node is different in terms
of the set of interactions $J_{ij}$. However, this difference can be pushed to the boundary conditions via the gauge transformation. And from the independence on boundary conditions in every state it follows that the distribution of fields is the same on every edge. 

In conclusion, the solution of the cavity equations for the
ground state of (\ref{Ham}) at zero magnetization are the same for
every $0\le\rho\le1$; this is true on any level of replica symmetry
breaking and, consequently, the ground state energy of (\ref{Ham})
is $\rho$-independent, as long as the graph of interactions is regular.

\section{Numerical evidence}

We use the extremal optimization (EO) heuristics \cite{BoettcherPercus01b,Dagstuhl04}
to find ground states of (\ref{Ham}) at zero magnetization for different
values of $\rho$. The EO heuristics has been used previously for
finding ground states on random graphs for graph bi-partitioning
($\rho=0$) \cite{Boettcher99a,Boettcher00,BoettcherPercus01,Percus08},
and for spin glasses ($\rho=1/2$) \cite{Boettcher03a,Boettcher03b}.
Thus, it is perfectly suited to approximate ground states over the
entire range of $0\leq\rho\leq1$. 

A detailed
study of the $\tau$-EO algorithm in its application to graph bi-partitioning and spin
glasses is already
provided in Refs.~\cite{Boettcher99a,BoettcherPercus01}, and we add only a
number of minor modifications here. EO considers each vertex of a
graph as an individual variable with its own fitness parameter. In the
graph bi-partitioning, or for any other $0<\rho\leq1$,  it assigns to each vertex $i$ a {}``fitness'' $\lambda_{i} = -b_{i}$,
where $b_{i}$ is the number of {}``bad'' (unsatisfied) edges connecting
$i$ to other vertices. 
At all times an ordered list is maintained, in the form of a permutation
$\Pi$ of the vertex labels $i$, such that 
$\lambda_{\Pi(1)}\leq\lambda_{\Pi(2)}\leq\ldots\leq\lambda_{\Pi(N)}$,
and $i=\Pi(k)$ is the label of the $k$-th ranked vertex in the list.
In its most elementary version, EO forces sequential updates of the
momentary worst variable $i=\Pi(1)$ at any update step, irrespective
of the outcome, inducing a cascade of adaptive reorderings in the list. Since all variables occupy an identical and
$O(1)$-sized state space, $\lambda_{i}=0,-1,\ldots,-r$, for
$r$-regular graphs, the list is highly degenerate and
maintaining order or selecting variables (with
fair tie-breaking rules) is done in $O(1)$ computations.

To define a local search of the configuration space, we must define a
{}``neighborhood'' for each configuration within this space. At zero
magnetization for all  $0\leq\rho\leq1$, as an
improvement over our previous implementation of EO for graph bi-partitioning, we proceed
here by allowing imbalanced partitions up to a margin of $\pm2$
vertices, independent of system size $N$. Then, we can pursue
single-variable updates as long as the resulting configuration remains
within the allowed imbalance. Valid ground states are only accepted, if the
current partition is perfectly balanced (although any $O(1)$ imbalance
for increasing $N$ should result in identical scaling behavior). In
this form, the single-flip neighborhood trivially generalizes to spin
glasses at freely fluctuating magnetization, for which we simply
ignore whether partitions remain balanced within the margins or not.

Much improved results are obtained with the following one-parameter
implementation \cite{BoettcherPercus01b}, called $\tau$-EO: An integer $1\leq
k\leq N$ is drawn from a probability distribution $P(k)\propto k^{-\tau}$, $1\leq k\leq N$,
on each update, for fixed $\tau$. Then, the vertex $i=\Pi(k)$ from the
rank-ordered list of fitnesses is selected for
an unconditional update. Over the course of a run (here with $t_{{\rm
    max}}=0.1N^{3}$ update steps), the costs of the configurations
explored varies widely, since each update can result in better or
worse fitnesses. The cost minimum of the \emph{best} configuration seen
during all runs for an instance is the output of the EO-algorithm. We
choose at least three uncorrelated restarts for each instance
here. The number of restarts is automatically adjusted for each
instance such that twice as many runs are undertaken than was
necessary to encounter  the putative ground state for the first
time. Only at larger system sizes and degree $r$, when more than 
10\% of instances require more than those three restarts, we initially
set the duration of each run to up to $t_{{\rm
    max}}=0.5N^{3}$ update steps.

Note that no scales to limit fluctuations are introduced into the
process, since the selection follows the scale-free power-law
distribution over ranks $P(k)$ and since all
moves are accepted. Instead of a global cost function, the
rank-ordered list of fitnesses provides the information about optimal
configurations. This information emerges in a self-organized manner,
merely by selecting with a bias against badly adapted variables,
rather than ever {}``breeding'' better ones. A theoretical analysis of
the optimal $\tau$-value is discussed at length in
Refs.~\cite{BoettcherPercus01,BoettcherPercus01b,eo_jam}, here, some initial
trials suggest  optimal values of $\tau=1.2-1.3$, which we have used
throughout.

Let us call $e_{{\rm GS}}(\rho,N)$ the ground state energy density,
averaged over graphs and disorder in interactions, of (\ref{Ham})
at magnetization fixed to zero, with $\rho$ being the fraction of
anti-ferromagnetic edges and $N$ the graph size. Denote by $\tilde{e}_{{\rm GS}}(\rho,N)$
the same quantity at an arbitrary magnetization. We have obtained
$e_{{\rm GS}}(\rho,N)$ on random regular graphs of degrees $r$ between
$3$ and $10$, and graph sizes between $N=32$ and $1024$. Statistical
errors of our averages have been kept small by generating a large
number of instances for each $N$ and $r$, typically $n_{I}\approx10^{6}$
for $N\leq200$, $n_{I}\approx10^{5}$ for $N\geq256$.

All our data are indeed consistent with the conjecture that on sparse random regular graphs  the ground state energy of (\ref{Ham}) at zero magnetization is $\rho$-independent in the thermodynamic limit, $N\to\infty$. However, we also observe that the finite-size
correction are $\rho$-dependent. This can be understood intuitively
on the very particular case of random 2-regular graphs. A random two-regular 
graph is basically a set of cycles of length $\sim\log{N}$.
Hence, whereas the bisection width is at most $2$ edges, the number
of edges minus the max-cut size is of order $N/\log{N}$ (a unit cost
for every other cycle in the graph). For $r\ge3$, the finite size
correction are not that large, but they are quite different for different
values of $\rho$.

\begin{table}[!ht]
\begin{centering}
\begin{tabular}{|c|c|c|c|c|}
\hline 
$r$ & $e_{{\rm 1RSB}}$ & $\tilde e_{{\rm GS}}\left(\rho=\frac{1}{2}\right)$ & $e_{{\rm GS}}\left(\rho=0\right)$ & $\omega\left(\rho=0\right)$\tabularnewline
\hline 
3  & -1.27231  & -1.2716(1)  & -1.2704(2) & 0.89\tabularnewline
\hline 
4  & -1.47295  & -1.472(1)  & -1.469(1) & 0.92\tabularnewline
\hline 
5  & -1.67520  & -1.673(1)  & -1.6717(5) & 0.86\tabularnewline
\hline 
6  & -1.82917  & -1.826(1)  & -1.824(1) & 0.87\tabularnewline
\hline 
7  & -1.99566  & -1.990(3)  & -1.990(1) & 0.85\tabularnewline
\hline 
8  & -2.12681  & -2.121(1)  & -2.120(2) & 0.87\tabularnewline
\hline 
9  & -2.27093  & -2.2645(5)  & -2.263(3) & 0.85\tabularnewline
\hline 
10  & -2.38769  & -2.378(3)  & -2.379(4) & 0.86\tabularnewline
\hline
\end{tabular}
\par\end{centering}
\caption{\label{tab:alldata} Asymptotic ground state energy per spin for different
values of the graph degree $r$. The second column, $e_{{\rm 1RSB}}$,
present $\rho$-independent one-step replica symmetry breaking results \cite{MezardParisi03}.
The third column, $e_{{\rm GS}}(1/2)$, contains the numerical values
of the extrapolated ground state energy for the spin glass\cite{Boettcher03a}, $\rho=1/2$.
The fourth and fifth column give  ground state energies $e_{{\rm GS}}(0)$ and  scaling coefficients $\omega(0)$
for the graph bisection problem obtained from infinite graph-size
extrapolations according to (\ref{fit}), as shown in Fig.~\ref{fig:ener_extra}.
With minor exceptions for the smallest degrees, $e_{{\rm GS}}(1/2)$
and $e_{{\rm GS}}(0)$ are the same within error bars. Moreover,
lacking a theoretical justification for Eq.~(\ref{fit}), the effective
error bars of the fitted values are larger than denoted in the table.}
\end{table}

\begin{figure}[!ht]
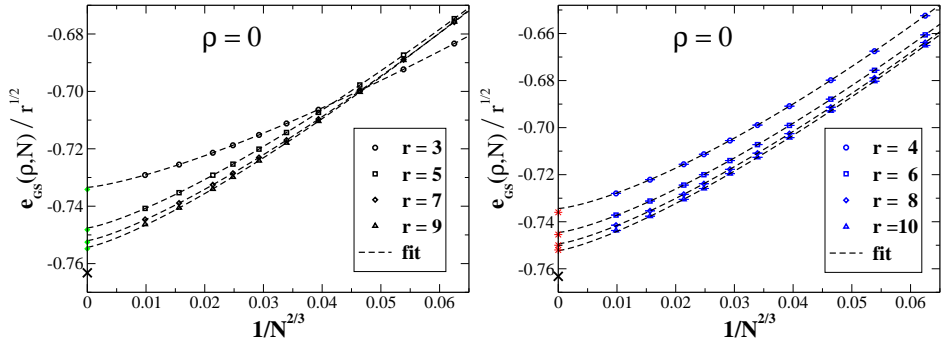

\includegraphics[width=0.47\linewidth]{extraGBPodd} \includegraphics[width=0.47\linewidth]{extraGBPeven}
\caption{\label{fig:ener_extra}Plot of the rescaled average
ground-state energy densities, $e_{{\rm GS}}(N)/\sqrt{r}$, for the
bi-partitioning of random regular graph of degree $r$ as a function
of $1/N^{2/3}$. Errors are smaller than symbol sizes and
are omitted for clarity. Non-linear fits to each data set according
to Eq.~(\ref{fit}) are indicated by dashed lines. Symbols on the
ordinate (green diamond for odd $r$, red star for even $r$) mark
the extrapolated values from Table \ref{tab:alldata} (rescaled by
$1/\sqrt{r}$) of the corresponding spin glass ground states 
from Ref.~\cite{Boettcher03a}. For large even and odd degree $r$,
the extrapolated values approach the ground state energy of the Sherrington-Kirkpatrick
model $U_{{\rm SK}}$ (black cross). The data are consistent with
the conjecture that the spin glass asymptotic ground states are equal
to the graph bi-partitioning ones.}
\end{figure}

In Tab.~\ref{tab:alldata}, we compare the asymptotic ground state energy
densities for different values of graph degree $r$. The second column
in this table presents the $\rho$-independent one-step replica symmetry breaking  results for the
ground state energy of (\ref{Ham}) at zero magnetization, computed
with the formalism developed in Ref.~\cite{MezardParisi03}. Note
that the exact value for the ground state would be provided by the
full-step replica symmetry breaking scheme which would give slightly
larger values. The data in the third column are taken from Ref.~\cite{Boettcher03a}.
In this case the magnetization was not fixed to be exactly zero on
every instance, but it is zero in density in the thermodynamic limit.
The data in \cite{Boettcher03a,BouchaudKrzakala03} are consistent with a power law scaling
\begin{equation}
e_{{\rm GS}}(N)=e_{{\rm GS}}+aN^{-\omega}\,.\label{fit}\end{equation}
with the value of the exponent $\omega=2/3$ for all $r$. The finite-size
scaling for the graph bisection is clearly not consistent with $\omega=2/3$,
as is illustrated in Fig.~\ref{fig:ener_extra} separately for odd
and even values of $r$\footnote{Note, however, that due to possibly strong higher order corrections to (\ref{fit}) the values of $\omega$ in Table~\ref{tab:alldata} may be skewed.}. As was noted in Refs.~\cite{Boettcher03a,Boettcher03b},
ground state energies are strongly affected by the existence or absence
of "free spins" on graphs with purely even or odd degrees, respectively.

\begin{figure}[!ht]
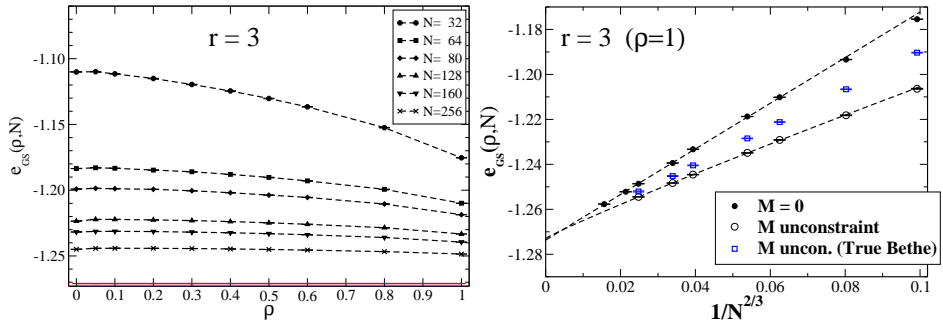

\begin{center}
\vspace{0.5cm}
\includegraphics[width=0.47\linewidth]{BondMixPhase}
\includegraphics[width=0.47\linewidth]{ener_extraAF100}
\end{center}
\caption{\label{fig:ener_rho}\label{fig:ener_AF} Left: The average ground state energy
density for finite-size 3-regular graphs as a function of the density
of anti-ferromagnetic bonds, $\rho$. The red dashed line marks the
value extrapolated for $\rho=0$ or $1/2$ from Tab.~\ref{tab:alldata}.
The data are consistent with the conjecture that the ground state
energy is asymptotically independent of $\rho$. Right:  Plot of average ground state energies
for the purely anti-ferromagnetic spin model ($\rho=1$) on 3-regular
graphs as a function of system size, both, at fixed ($M=0$) and at
unconstraint magnetization. The constraint case appears to extrapolate
well linearly on a $N^{-2/3}$ scale without any transient,
while the unconstraint case shows a small deviation of about 5\% when
all data is fitted according to Eq.~(\ref{fit}). With blue squares, we also
plot data for the same problem with unconstraint $M$ but disallowing
multiple edges between two vertices. All other data was obtained
allowing graphs with such edges. }
\end{figure}

On the left hand side of Fig.~\ref{fig:ener_rho}, we study numerically the dependence of
average ground-state energies as a function of $\rho$ at finite sizes
on 3-regular graphs only. While there are significant differences
in the magnitude of corrections -- even on average -- for the smaller sizes,
finite-size behavior soon becomes virtually independent of $\rho$
and appears to converge towards the thermodynamic values found consistently
at $\rho=0$ and $1/2$ listed in Tab.~\ref{tab:alldata}. 

On the right hand side of Fig.~\ref{fig:ener_AF}, we address the question on how finite-size corrections are impacted by the constraint on magnetization being fixed to zero. We compare the average ground state energy of the anti-ferromagnet ($\rho=1$) with magnetization strictly zero and with no constraint on magnetization. Clearly, in the unconstrained case the
distribution of ground-state magnetizations is symmetrical above $\rho_{c}$,
with vanishing fluctuations in the thermodynamic limit. Hence, the
average magnetization must be zero. Yet, at finite size, it seems
conceivable that fixing the magnetization makes otherwise lower-energy
states, i.~e. ground states of the corresponding unconstraint model,
unattainable, which may shift the ground state energies of the constraint
system upwards. Indeed, as Fig.~\ref{fig:ener_AF} demonstrates, while
indistinguishable thermodynamically, average energies increase by large
amounts relative to the unconstraint case especially for small sizes.
But the actual finite-size corrections seems to be compatible with
$\omega=2/3$ in both cases. In our simulations, we generate $r$-regular graphs in such
a way that multiple edges between identical vertices are not forbidden
in their random assignment. Such multi-linked vertices have a probability
of $\sim1/N$ and, although noticeable at small size, do not affect
any asymptotic scaling, see Fig.~\ref{fig:ener_AF}. Forbidding such edges makes it difficult to
generate valid graphs especially at larger $r$ and small sizes.

\section{Possible generalizations}

As we have argued above, the relation between max-cut and bisection
width does not generalize (at least not straightforwardly) to the
case of non-zero magnetization (not equally-sized groups) nor to the
case of non-regular graphs. However, it does generalize to finite
temperature properties of the Hamiltonian (\ref{Ham}) at zero magnetization.

The relation between max-cut being the number of edges minus the bisection
width also generalizes to the case of hyper-graphs. In statistical
physics, corresponding models are know as models with $p$-spin interactions,
in computer science as the XOR-SAT problem (boolean constraint satisfaction
problem consisting of sets of linear equations).

An alluring but not (fully) valid generalization to discuss
is the case of Potts spins $s_{i}=1,\dots,k$. The max-cut problem
is then replaced by max-$k$-coloring problem and the bisection by
 $k$-partitioning. According to a result by Kanter and Sompolinsky
\cite{KanterSompolinsky87}, analogous to the one of \cite{FuAnderson86},
in dense graphs the two problems are related. Be $pN$ the degree
of the graph, and $k$ the number of colors, then according to Ref.~\cite{KanterSompolinsky87}
the maximum number of non-monochromatic edges is \begin{equation}
|{\rm MaxCol}|=\frac{p}{2}N^{2}\Big(1-\frac{1}{k}\Big)+N^{\frac{3}{2}}\frac{|U(k)|}{k}\sqrt{p(1-p)},\label{col}\end{equation}
 whereas the minimal number of edges between groups in the best balanced
$k$-partition is \begin{equation}
|k{\rm -part}|=\frac{p}{2}N^{2}\Big(1-\frac{1}{k}\Big)-N^{\frac{3}{2}}\frac{|U(k)|}{k}\sqrt{p(1-p)}.\label{par}\end{equation}
Here, $U(k)$ is in both the expressions for the ground state energy
of the fully-connected Potts model with $k$ colors \cite{GrossKanter85},
numerically given in Ref.~\cite{KanterSompolinsky87}, while in the 
large-$k$ limit it is $\lim_{k\to\infty}U(k)=\sqrt{(k\ln{k})}$. For $k=2$
the relations (\ref{col}-\ref{par}) reduce to the Fu and Anderson
result (\ref{FuA}).

Based on the analogy of the dense graph case, we would thus expect
a generalization for  sparse graphs, $p=c/N$, also for $k>2$.
However, for three and more colors there is no apparent relation between
the max-coloring and $k$-partitioning. Whereas there are some version
of the Potts glass equivalent to the coloring problem, see e.~g. Ref.~\cite{KrzakalaZdeborova07},
there is no obvious Gauge transform able to transfer the Potts ferromagnet
(allowing $1$ out of $q$ values) on the Potts anti-ferromagnet (allowing
$q-1$ out of $q$ values). This can be seen explicitly in the difference
between the replica symmetric equations for the two problems. In the warning-propagation sense \cite{BraunsteinMezard02}, the neutral warning
in max-coloring is created, if two colors have the same value of an
incoming field and the third one has a larger value. In contrast,  in partitioning
the third value needs to be smaller. Also, the expressions for the
replica symmetric energy are different, even after numerical evaluation.
The equivalence thus holds only asymptotically in the first two orders
of the degree of the graph (as suggested by the result in Ref.~\cite{KanterSompolinsky87}).
This underlines the exceptional nature of our main conjecture (\ref{con})
for $k=2$.

\section{Conclusion}

In this note we describe, explain, and support by numerical evidence a
conjecture that on sparse random regular graphs the ground state
energy value of the spin glass Hamiltonian (\ref{Ham}) at
magnetization fixed to zero does not depend on the fraction of
anti-ferromagnetic bonds. Although hints towards this conjecture can
be found in the existing literature, we state it as a clear
mathematical conjecture understandable for non-specialist in  spin
glass theory: In random regular graphs, the asymptotic size of the
max-cut equals the number of edges minus the minimal bisection
width. We also summarize necessary conditions and limitations of this
conjecture, in particular, that it does not generalize (at least not
in a way we could see) to non-regular graphs and non-zero values of
the magnetization. We also support the conjecture by extensive
numerical evaluations of the ground states. Finally, we are positive that this note will be useful for the mathematical and computer science community and that it will lead to a proof of this conjecture in the near future.

\paragraph*{Acknowledgment}

The authors thank Florent Krzakala, Cris Moore and Petr \v{S}ulc for very
useful discussions. SB acknowledges support from the Fulbright Kommission
and from the U. S. National Science Foundation through grant number
DMR-0812204.

\section*{References}

\bibliographystyle{unsrt}
\bibliography{myentries,/Users/stb/Boettcher}

\end{document}